\documentclass[english,aps, prb, twocolumn, superscriptaddress,showpacs,floatfix]{revtex4-1}
\usepackage{lmodern}
\usepackage[T1]{fontenc}
\usepackage[latin9]{inputenc}
\usepackage{textcomp}
\usepackage{graphicx}
\usepackage{bm}
\usepackage{verbatim}
\usepackage{booktabs}
\usepackage{amstext}
\usepackage{amsmath}
\usepackage{color}
\usepackage{upgreek}
\usepackage{multirow}
\usepackage{siunitx}
\usepackage{paralist}

\usepackage{printlen}

\makeatletter

\newcommand{\lyxmathsym}[1]{\ifmmode\begingroup\def\b@ld{bold}
  \text{\ifx\math@version\b@ld\bfseries\fi#1}\endgroup\else#1\fi}

\@ifundefined{textcolor}{}
{%
 \definecolor{BLACK}{gray}{0}
 \definecolor{WHITE}{gray}{1}
 \definecolor{RED}{rgb}{1,0,0}
 \definecolor{GREEN}{rgb}{0,1,0}
 \definecolor{BLUE}{rgb}{0,0,1}
 \definecolor{CYAN}{cmyk}{1,0,0,0}
 \definecolor{MAGENTA}{cmyk}{0,1,0,0}
 \definecolor{YELLOW}{cmyk}{0,0,1,0}
 \definecolor{darkgreen}{rgb}{0,0.6,0.4}
\definecolor{azzuri}{rgb}{0.2,0.2,0.7}
 }

\usepackage{prettyref}

\newcommand{\etal}{\textit{et al.}}

\newcommand{\col}[1]{{\color{BLACK} #1}}       

\newcommand{\chisq}{{\ensuremath{\chi^2}}}

\newcommand{\LFO}{{LuFe$_2$O$_4$}}	   
\newcommand{\Fetwo}{{Fe$^{2+}$}}	   
\newcommand{\Fethree}{{Fe$^{3+}$}}    

\newcommand{\E}{{\ensuremath{E}}}
\newcommand{\Q}{\ensuremath{\bm{Q}}}

\newcommand{\Ei}{{\ensuremath{E_{\rm i}}}}

\newcommand{\degree}{{\ensuremath{^{\circ}}}}
\newcommand{\degreeC}{{\ensuremath{^{\circ}{\rm C}}}}
\newcommand{\invAA}{{\AA\ensuremath{^{-1}}}}

\makeatother

\usepackage{babel}

\begin{document}

\graphicspath{{.}{Figures/}}

\title{Magnetic excitation spectrum of \LFO\ measured with inelastic neutron scattering}

\author{S. M. Gaw}
\email[]{s.gaw1@physics.ox.ac.uk}
\affiliation{Department of Physics, University of Oxford, Clarendon Laboratory,
Parks Road, Oxford, OX1 3PU, United Kingdom}

\author{H. J. Lewtas}
\altaffiliation{Current Address: BAE SYSTEMS, Advanced Technology Centre, Burcote Road, Towcester, NN12 7AR}
\affiliation{Department of Physics, University of Oxford, Clarendon Laboratory,
Parks Road, Oxford, OX1 3PU, United Kingdom}

\author{D. F. McMorrow}
\affiliation{London Centre for Nanotechnology and Department of Physics and Astronomy, University College London, London WC1E 6BT, United Kingdom}

\author{J. Kulda}
\affiliation{Institut Laue-Langevin, B.P. 156, 38042 Grenoble, France}

\author{R. A. Ewings}
\author{T. G. Perring}
\affiliation{ISIS Facility, Rutherford Appleton Laboratory, STFC, Chilton, Didcot,
Oxon, OX11 0QX, United Kingdom}

\author{R. A. M$^{\rm c}$Kinnon}
\author{G. Balakrishnan}
\affiliation{Department of Physics, University of Warwick, Coventry, CV4 7AL, United Kingdom}

\author{D. Prabhakaran}
\author{A. T. Boothroyd}
\email[]{a.boothroyd@physics.ox.ac.uk}
\affiliation{Department of Physics, University of Oxford, Clarendon Laboratory,
Parks Road, Oxford, OX1 3PU, United Kingdom}

\begin{abstract}
We report neutron inelastic scattering measurements and analysis of the spectrum of magnons propagating within the Fe$_2$O$_4$ bilayers of \LFO. The observed spectrum is consistent with six magnetic modes and a single prominent gap, which is compatible with a single bilayer magnetic unit cell containing six spins. We model the magnon dispersion by linear spin-wave theory and find very good agreement with the domain-averaged spectrum of a spin--charge bilayer superstructure comprising one Fe$^{3+}$-rich monolayer and one Fe$^{2+}$-rich monolayer. These findings indicate the existence of polar bilayers in \LFO, contrary to recent studies that advocate a charge-segregated non-polar bilayer model. Weak scattering observed below the magnon gap suggests that a fraction of the bilayers contain other combinations of charged monolayers not included in the model. Refined values for the dominant exchange interactions are reported.
\end{abstract}
\maketitle

\section{\label{sec:Introduction} Introduction}


\LFO\ is a complex and controversial material exhibiting a variety of behaviors and ordering phenomena.\cite{Ikeda2008,AngstPhysStatusSolidi2013} Recent interest in \LFO\ stems from a proposal that it exhibits a novel type of ferroelectricity driven by charge ordering (CO) of Fe$^{2+}$ and Fe$^{3+}$ oxidation states (`electronic ferroelectricity') which also couples to magnetic order.\cite{Ikeda2005} In addition, \LFO\ was reported to have a giant dielectric response which strongly couples to an applied magnetic field.\cite{Subramanian2006} However, subsequent studies have shown that the large dielectric constant can be explained through non-intrinsic effects.\cite{Ren2011, Niermann2012, Ruff2012} Furthermore, the pattern of charge order originally proposed to explain electronic ferroelectricity in \LFO\ has recently been called into question.\cite{deGroot2012May, Ruff2012} Alternative charge segregation\cite{deGroot2012May} and anti-polar\cite{Xu2010} CO models have been proposed.

The goal of the present study was to interrogate the ordered ground state of \LFO\ through the magnetic excitation spectrum. In magnetically ordered systems the spectrum of magnetic excitations, or magnons, depends on the nature of the magnetic order and on the exchange interactions that stabilise it. In \LFO\, these properties will in turn depend on the CO state. Inelastic neutron scattering measurements on a powder sample of \LFO\ revealed a magnon gap of approximately 8\,meV,\cite{Bourgeois2012Jul} but up to now no measurements on single crystals capable of resolving the magnon dispersion relations have been published.


\begin{figure}[h!]
\centering
\includegraphics[scale=0.37, trim=5mm 25mm 5mm 25mm ]{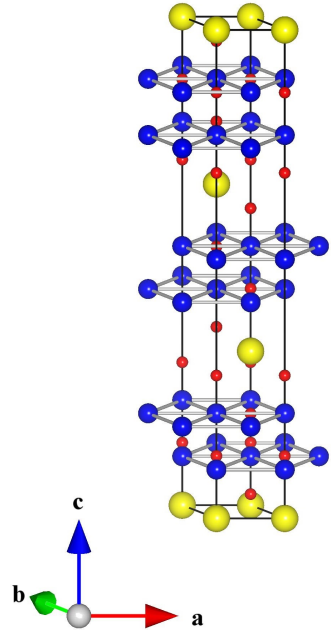}
\caption{\label{fig:LFO_crystal_structure} The crystal structure of \LFO, space group $R\bar{3}m$. Lu, Fe, and O atoms are depicted as yellow, blue and red spheres respectively. The grey bonds indicate the triangular coordination of Fe atoms in the monolayers. These are stacked to form bilayers, which are in turn separated by LuO$_2$ layers. The $R$-centred hexagonal unit cell is shown.}
\end{figure}

The room temperature crystal structure of \LFO, shown in Fig.~\ref{fig:LFO_crystal_structure}, is described by the rhombohedral space group $R\bar{3}m$.\cite{Isobe1990} The lattice parameters in the hexagonal setting are $a=b=3.44$\,\AA, $c=25.3$\,\AA\ with inter-axis angles $\alpha=\beta=90$\degree, $\gamma=120$\degree. The structure can be considered as a stacking of Fe$_2$O$_4$ bilayers along the $c$ axis, each bilayer consisting of two monolayers on which the Fe atoms form a triangular lattice.  There are three bilayers per unit cell. The complex ordering phenomena exhibited by \LFO\ arise because in the ideal structure all Fe sites are equivalent with average valence Fe$^{2.5+}$. However, already at room temperature there is near-perfect charge disproportionation into \Fetwo\ and \Fethree.\cite{Tanaka1993,Ikeda2005,Ko2009,Mulders2009,deGroot2012May} The distribution of equal amounts of \Fetwo\ and \Fethree\ on the triangular layers creates frustration which influences the CO and magnetic order.

\begin{figure}
\includegraphics[width=\columnwidth]{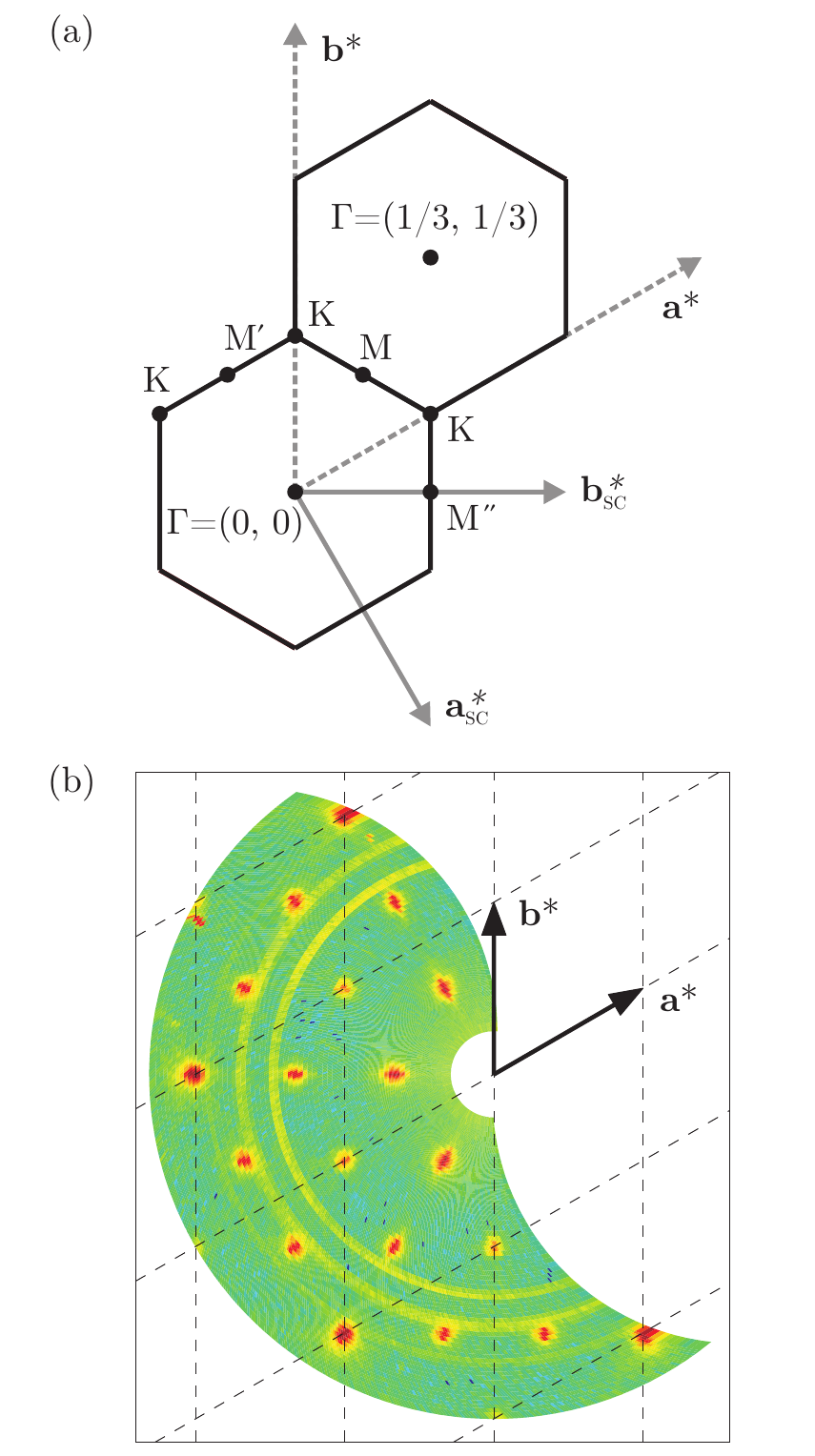}
\caption{\label{fig:LFO_recip_latt}(Color online) (a) Reciprocal lattice and Wigner--Seitz cells of the charge- and magnetically-ordered superstructure of \LFO. The direction of the reciprocal lattice vectors for the crystallographic unit cell (${\bf a}^{\ast}$ and ${\bf b}^{\ast}$) and supercell (${\bf a}^{\ast}_{\rm SC}$ and ${\bf b}^{\ast}_{\rm SC}$) are shown as dashed and solid grey arrows, respectively. The high symmetry points of the Brillouin zone are labelled. (b) Neutron diffraction pattern in the $(H,K,0)$ plane of Sample 1 measured at a temperature of 10\,K on IN20. The ${\bf a}^{\ast}$ and ${\bf b}^{\ast}$ reciprocal lattice vectors (black arrows) are shown together with the reciprocal lattice grid (dashed lines). Magnetic peaks are visible at the ${\bf q}_{\rm m}=(1/3,1/3)$ reduced wave vector and equivalent positions.}
\end{figure}

Charge order of \Fetwo\ and \Fethree\ is detectable in \LFO\ below $\sim 500$\,K. The CO is initially quasi-two-dimensional (2D), but develops into three-dimensional (3D) long-range order on cooling below about 320\,K.\cite{Yamada2000} The charge ordering is identified by superstructure peaks in diffraction measurements with in-plane wave vectors very close to ${\bf q}_{\rm CO}=(1/3,1/3)$ and equivalent positions --- here and elsewhere in this paper all wave vectors are expressed in reciprocal lattice units of the hexagonal lattice, for which $a^{\ast} = b^{\ast} = 2/\surd{3} \times 2\pi/a$ (see Fig.~\ref{fig:LFO_recip_latt}). In reality, a small shift $(\delta,\delta)$, where $\delta \sim 0.003$, away from the commensurate superstructure positions is observed, which is most likely caused by regular discommensurations or anti-phase boundaries in the ideal commensurate charge order.\cite{AngstPhysStatusSolidi2013} Our neutron spectroscopy measurements are not sensitive to this small shift and we shall neglect it hereafter.

\begin{figure}
\includegraphics[width=\columnwidth]{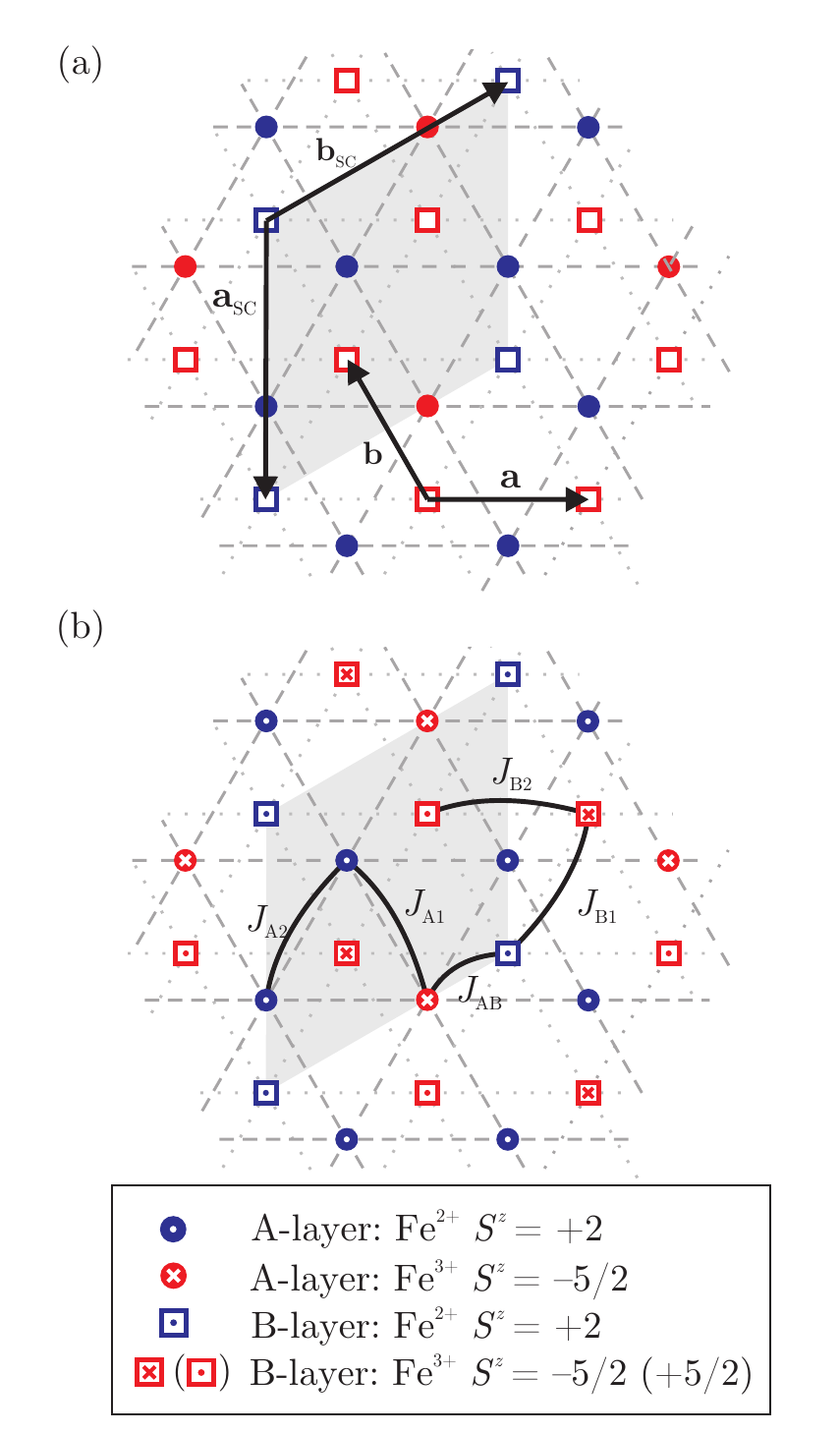}
\caption{\label{fig:LFO_charge_mag_struct}(Color online) (a) Projection down the $c$ axis of a charge-ordered AB-bilayer. The lattice vectors ${\bf a}$ and ${\bf b}$ and superlattice vectors ${\bf a}_{\rm SC}$ and ${\bf b}_{\rm SC}$ are shown as black arrows. The ions in the \Fetwo-rich A monolayers (filled circles) and \Fethree-rich B monolayers (open squares) are projected on top of one another. Oxygen atoms have been omitted for clarity. Blue and red symbols denote \Fetwo\ and \Fethree\ ions, respectively. The $\sqrt{3}\times\sqrt{3}$ CO supercell is depicted by the grey diamond. (b) Spin structure of a ferromagnetically-coupled AB-bilayer (see main text). The spins point along the $c$-axis, either into (cross) or out of (dot) the page. The five exchange pathways considered in the minimal spin-wave model are labelled.}
\end{figure}

 One way to overcome the charge frustration on the triangular layers and reproduce the observed ${\bf q}_{\rm CO}$ is through the formation of \Fetwo-rich monolayers (with 2:1 ratio of \Fetwo:\,\Fethree, designated here an A-layer) and \Fethree-rich monolayers (1:2 ionic ratio, B-layer) in equal proportions.\cite{Yamada2000, Ikeda2005} The CO in an A-layer is a honeycomb network of \Fetwo\ with \Fethree\ at the centre of each \Fetwo\ hexagon, while the B-layer has just the opposite arrangement of \Fetwo\ and \Fethree. Figure~\ref{fig:LFO_charge_mag_struct}\col{(a)} shows a projection down the $c$-axis of a bilayer made from an A- and a B-layer. The CO of each monolayer (and the resulting bilayer) is described by an enlarged $\sqrt{3}\times\sqrt{3}$ supercell (SC), shown by the grey diamond in Fig.~\ref{fig:LFO_charge_mag_struct}\col{(a)}. Such an AB bilayer has a net electric dipole moment, which prompted the original proposal for electronic ferroelectricity in \LFO.\cite{Ikeda2005}  The AB bilayer model was found to be consistent with \emph{ab initio} calculations,\cite{Xiang2007,Angst2008} and diffraction data was initially interpreted in terms of an antiferroelectric bilayer stacking (AB-BA) with defects in the form of ferroelectric short-range correlations (AB-AB) between neighbouring bilayers.\cite{Angst2008} Later, however, a full crystal structure refinement was carried out against single crystal X-ray diffraction data recorded at $210$\,K, and a model emerged with an AA-BB stacking corresponding to alternately charged bilayers.\cite{deGroot2012May} 


\LFO\ orders magnetically below $\rm{T_N}\approx 240$~K.\cite{Iida1993, Christianson2008} The \Fetwo\ ($3d^6$) and \Fethree\ ($3d^5$) ions both carry an ordered magnetic moment governed largely by their spin states of $S=2$ and $S=5/2$, respectively. The sizes of the ordered moments are estimated to be 4.8\,$\mu_{\rm B}$ (\Fetwo) and 5\,$\mu_{\rm B}$ (\Fethree).\cite{Ko2009} The \Fetwo\ moment is greater than the spin-only value of 4\,$\mu_{\rm B}$ because it has a significant orbital component due to the combined effects of spin-orbit coupling and the trigonal bipyramidal crystal field,\cite{Ko2009, Kuepper2009} which causes the moment to have an easy axis along the $c$ axis.\cite{Iida1993} The orbital moment is almost fully quenched in \Fethree\ leading to isotropic magnetism, but magnetic coupling to the Ising-like \Fetwo\ effective spins causes the ordered moments on all Fe sites to point either parallel or antiparallel to the $c$ axis.

The spin structure on the triangular layers has been investigated by M\"{o}ssbauer spectroscopy,\cite{Tanaka1993} neutron diffraction,\cite{Iida1993,Christianson2008, deGroot2012Jan} X-ray spectroscopy,\cite{Ko2009,Kuepper2009} and X-ray resonant magnetic scattering.\cite{deGroot2012Jan} The measurements show that the monolayers are ferrimagnetic (fM) with a $\uparrow\uparrow\downarrow$ configuration, and that the in-plane magnetic ordering wave vector of \LFO\ is the same as that of the individual monolayers and is ${\bf q}_{\rm m} = {\bf q}_{\rm CO} = (1/3,1/3)$. The magnetic superstructure in \LFO\ is illustrated in Fig.~\ref{fig:LFO_recip_latt}(b), which displays the neutron diffraction intensity in the $(H,K,0)$ plane recorded at 10\,K. The range of the magnetic correlations along the $c$ axis appears to be sample-dependent, with quasi-2D order observed in some samples,\cite{Iida1993} but full 3D order observed at temperatures just below $T_{\rm N}$ in others.\cite{Christianson2008,deGroot2012Jan}

The weight of evidence suggests the following spin arrangements on the A and B layers: (i) on the A-layer, the majority \Fetwo\ spins are all parallel to one another forming a ferromagnetic honeycomb net, and the minority \Fethree\ spins are all aligned in the opposite direction; (ii) on the B-layer, the majority \Fethree\ spins form an antiferromagnetic honeycomb net with each spin antiparallel to its nearest \Fethree\ neighbour, while the minority \Fetwo\ spins are all parallel to one another creating the fM moment.

\begin{figure}
\centering
\includegraphics[width=\columnwidth]{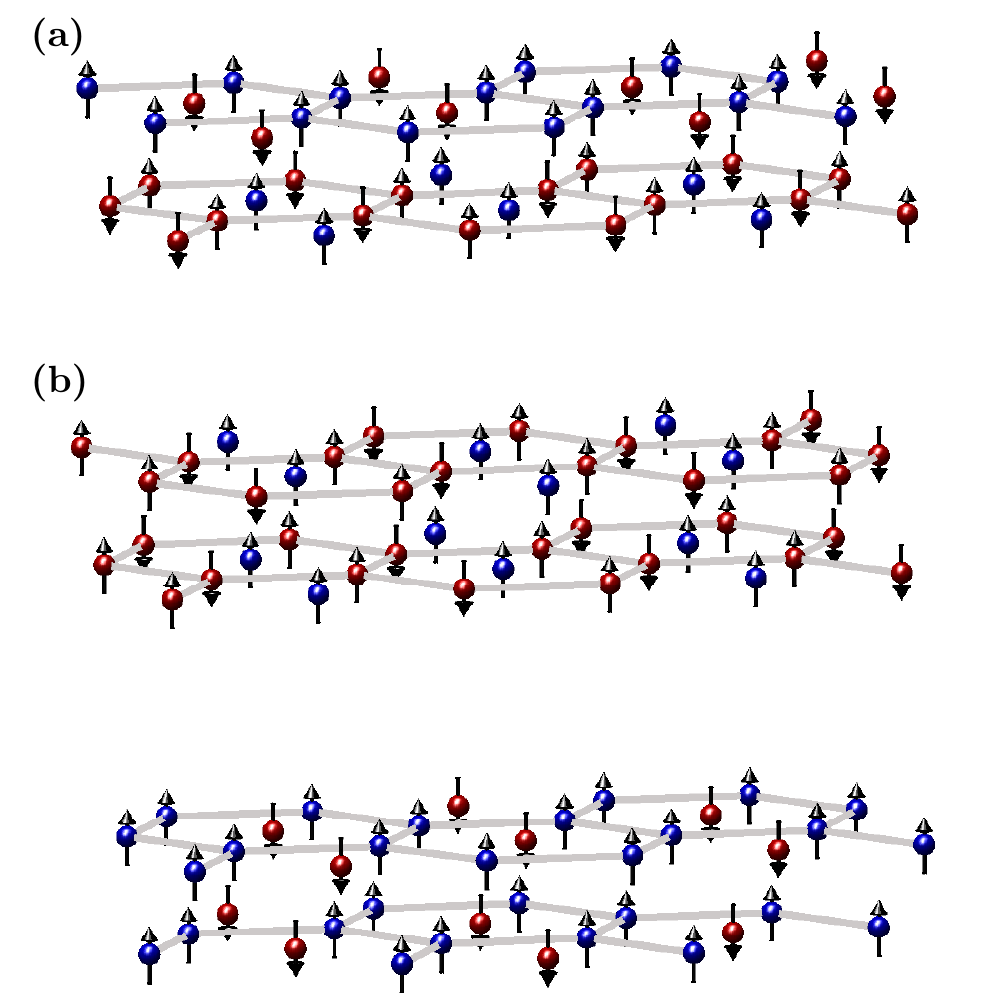}
\caption{\label{fig:LFO_3D_struct_comparison}(Color online) Proposals for the charge- and magnetically-ordered bilayer superstructures in \LFO. (a) The AB bilayer proposed in Ref.~\onlinecite{Ko2009}. (b) The AA-BB bilayers proposed in Ref.~\onlinecite{deGroot2012Jan}. Blue (red) spheres denote the \Fetwo\ (\Fethree) ions. Arrows denote the ordered magnetic moments. Grey lines show the \Fetwo--\Fetwo\ bonds on the A-layer and the \Fethree--\Fethree\ bonds on the B-layer. This highlights the honeycomb structure formed by the majority ions on each layer. The minority ion is found in the middle of each hexagon.}
\end{figure}

Given the monolayer spin structures just described, there are several possibilities for the bilayer spin structure.\cite{AngstPhysStatusSolidi2013} Magnetisation data show that the fM moments on adjacent monolayers in a bilayer must be parallel to one another to account for the observed saturated moment of 2.9\,$\mu_{\rm B}$ per \LFO\ formula unit,\cite{Iida1993} although a refinement of the magnetic structure against neutron powder diffraction suggested a temperature-dependent phase mixture of parallel (ferromagnetic, FM bilayer) and antiparallel (antiferromagnetic, AFM bilayer) monolayer moments.\cite{Bourgeois2012Jul} Analysis of X-ray spectroscopy data based on an AB bilayer indicated that all \Fetwo\ spins in the bilayer are parallel to one another.\cite{Ko2009,Kuepper2009}  This arrangement is shown in projection in Fig.~\ref{fig:LFO_charge_mag_struct}\col{(b)}, and depicted in 3D in \ref{fig:LFO_3D_struct_comparison}\col{(a)}. The AA-BB bilayer model proposed in Ref.~\onlinecite{deGroot2012Jan} has the magnetic structure shown in Fig.~\ref{fig:LFO_3D_struct_comparison}\col{(b)}. If one simply considers the size and orientation of the moments then the AA-BB model has almost the same magnetic structure as an AB-BA model, since the ordered moments on \Fetwo\ and \Fethree\ are almost the same. However, the single-ion magnetic anisotropy and two-ion magnetic couplings depend strongly on the Fe valence states, so the two models will have different magnon spectra, as discussed below. The weak inter-bilayer coupling leads to antiparallel stacking of adjacent fM bilayers, but there is evidence for considerable (sample-dependent) disorder in the stacking along the $c$ axis, which is likely the reason why quasi-2D magnetic order and spin-glass behavior is often observed.\cite{Wu2008, Wang2009,deGroot2012Jan, Bourgeois2012Jul}

 In this work we performed inelastic neutron scattering (INS) measurements of the complete magnon spectrum of \LFO. The observed spectrum is consistent with six distinct magnetic modes, which suggests that the majority of the sample has the AB-type bilayer magnetic ordering. The data allows the possibility of a small proportion of AA and BB bilayers. A minimal spin-wave model comprising five exchange interactions and single-ion anisotropy is shown to give a very good description of the observed magnon dispersion.

\section{\label{sec:Experimental Details} Experimental Details}

Two single crystal samples, prepared in Oxford and Warwick (denoted Samples 1 and 2), were grown via the same basic procedure. High purity ($>99.999$\,\%) Lu$_2$O$_3$ and Fe$_2$O$_3$ powders were mixed in the stoichiometric ratio of \LFO. This mixture was sintered at 1200\degreeC\ for 12\,h under a flowing CO/CO$_2$ atmosphere (in a 25/75\% ratio for Sample 1, 17/83\% for Sample 2). The mixture was re-ground and heat treated a second time at 1200\degreeC\ for 24\,h in an Ar atmosphere. The powder was subsequently pressed into a rod (8\,mm diameter, 100\,mm length) and sintered for 12\,h at 1200\degreeC\ for Sample 1 and 1250\degreeC\ for Sample 2 in a CO/CO$_2$ atmosphere (Sample 1: 30/70\%, Sample 2: 17/83\%). The crystal growth was performed with optical floating-zone furnaces (Crystal Systems Inc.) in a flowing CO/CO$_2$ atmosphere (Sample 1: 5/95\%, Sample 2: 17/83\%). Feed and seed rods counter-rotated in this setup at 30\,rpm, with the growth proceeding at 1--2\,mm/h and 0.5--1\,mm/h for Sample 1 and 2 respectively. The initial growth was performed with polycrystalline seed rods, but subsequent growths used a cleaved single crystal as a seed.   The growths of \LFO\ typically yielded multigrain samples. The two high-quality single-grain crystals used here were cleaved from such growths. The masses of the crystals were 0.35\,g (Sample 1) and 2.47\,g (Sample 2). X-ray and neutron Laue diffraction confirmed that no secondary grains remained in either sample. Magnetisation measurements were performed with a superconducting quantum interference device (SQUID) magnetometer (Quantum Design).

The neutron triple-axis spectrometers (TAS) IN20 and IN8 at the ILL, Grenoble, were used to measure the in-plane diffraction pattern and low energy excitations of Sample 1. On both instruments the crystal was aligned with $(H,K,0)$ as the horizontal scattering plane. The IN8 measurement employed double-focussing Si $(111)$ and double-focussing Cu $(200)$ monochromators, in combination with a double-focussing pyrolytic graphite $(002)$ analyser, to measure neutrons scattered with fixed final wave vectors of either $k_{\rm f}=2.662$ or 4.1\,\invAA. Measurements were made at a temperature of 1.5\,K using a standard helium cryostat. On IN20, we used the FlatCone multiplexed secondary spectrometer with lifting capability to measure scattering in the $(H,K,L)$ plane for fixed $L$ values in the range $0 \leq L \leq 2$. Again, a double-focussing Si $(111)$ monochromator was used. An array of Si $(111)$ analysers selected neutrons with final wave vector $k_{\rm f}=3.0$\,\invAA. Data were collected on FlatCone at 10\,K.

The larger mass of Sample 2 allowed the use of time-of-flight (TOF) spectroscopy to survey the excitation spectrum across a wide range of energy \E\ and wave vector ${\bf Q}=(H\times 2\pi/a,K\times 2\pi/a,L\times 2\pi/c)$. The measurements were made on the MAPS spectrometer at the ISIS Facility, UK. The crystal was aligned with the $c$ axis parallel to the incident neutron wave vector ${\bf k}_{\rm i}$. In this configuration, the $L$ component of ${\bf Q}$ varies strongly with \E. However, scans parallel to $(0,0,L)$ made on Crystal 1 did not reveal any measurable dispersion of the lowest measurable magnetic excitation, which confirms that the inter-bilayer coupling is very weak in \LFO. Hence, the measured intensity corresponds to the single-bilayer spectrum and can be described by the 2D wave vector ${\bf Q}=(H,K)$.  The variation in intensity with $L$ due to the bilayer magnetic structure factor, the magnetic form factor and the orientation factor (see Eq.~\ref{eq:S(Q,E)}),  was included in the model simulations. Preliminary high-flux measurements were made using incident energies of $E_{\rm i}=60,~80,~120,~150$ and 200\,meV to identify the total bandwidth of the spectrum. The energy resolution broadening as defined by the full width at half maximum (FWHM) at $E=0$\,meV was approximately 6\% of $E_{\rm i}$. This broadening decreases with increasing $E$. Higher resolution configurations were used to repeat the measurements with $E_{\rm i}=60$ and 80\,meV, where the FWHM was reduced to approximately 4\% of $E_{\rm i}$ at $E=0$\,meV. The sample was mounted in a closed-cycle refrigerator, and data were recorded at a temperature of approximately 7\,K.

The intensity of magnetic scattering is described by the partial differential cross-section, which is expressed in the dipole approximation as,\cite{Squires}
\begin{equation}
\frac{\partial ^{2}\sigma}{\partial\Omega \partial E}=\frac{k_{\rm f}}{k_{\rm i}}\Big(\frac{\gamma r_0}{2}\Big)^{2}|f(Q)|^2 \sum_{\alpha}(1-\hat{Q}_{\alpha}^2) S^{\alpha\alpha}({\bf Q},E).
\label{eq:S(Q,E)}
\end{equation}
In this equation, $k_{\rm i}$ and $k_{\rm f}$ are the incident and final neutron wave vectors, $(\gamma r_0/2)^2=72.8$ mb, $f(Q)$ is the magnetic form factor (assumed the same for \Fetwo\ and \Fethree) and $(1-\hat{Q}_{\alpha}^2)$ is the orientation factor. We have neglected the Debye--Waller factor, which is close to unity at low temperatures. $\hat{Q}_{\alpha}=Q_{\alpha}/Q$ is the $\alpha$ component of the unit vector parallel to $\bf Q$.

The scattering function $S^{\alpha\alpha}({\bf Q},E)$ describes magnetic correlations between the $\alpha$ components of the magnetic moments ($\alpha= x, y, z$ in Cartesian coordinates). When the orbital angular momentum is quenched, the scattering function may be written
\begin{equation}
S^{\alpha\alpha}({\bf Q},E)=g^2_{\alpha}\sum_{j}|\langle j|S^{\alpha}({\bf Q})|0\rangle |^{2} \delta(E - E_j({\bf Q})),
\label{eq:Saa}
\end{equation}
where $g_{\alpha}$ is a diagonal component of the $g$-factor tensor, $S^{\alpha}({\bf Q})$ is a component of the spin dynamical structure factor, $\left|0\right\rangle$ is the magnon vacuum state and $\left|j\right\rangle $ is an excited magnon state with energy $E_{j}$.

The measured scattering intensity was corrected for the factor of $k_{\rm f}/k_{\rm i}$ in Eq.~\ref{eq:S(Q,E)}, and for time-independent backgrounds. Measurements were performed on a standard vanadium sample to allow correction for the counting efficiencies of individual detectors and to convert intensities to an absolute scale.

\section{\label{sec:Results} Results}

\subsection{Magnetic Characterization}

\begin{figure}
\centering
\includegraphics[width=\columnwidth]{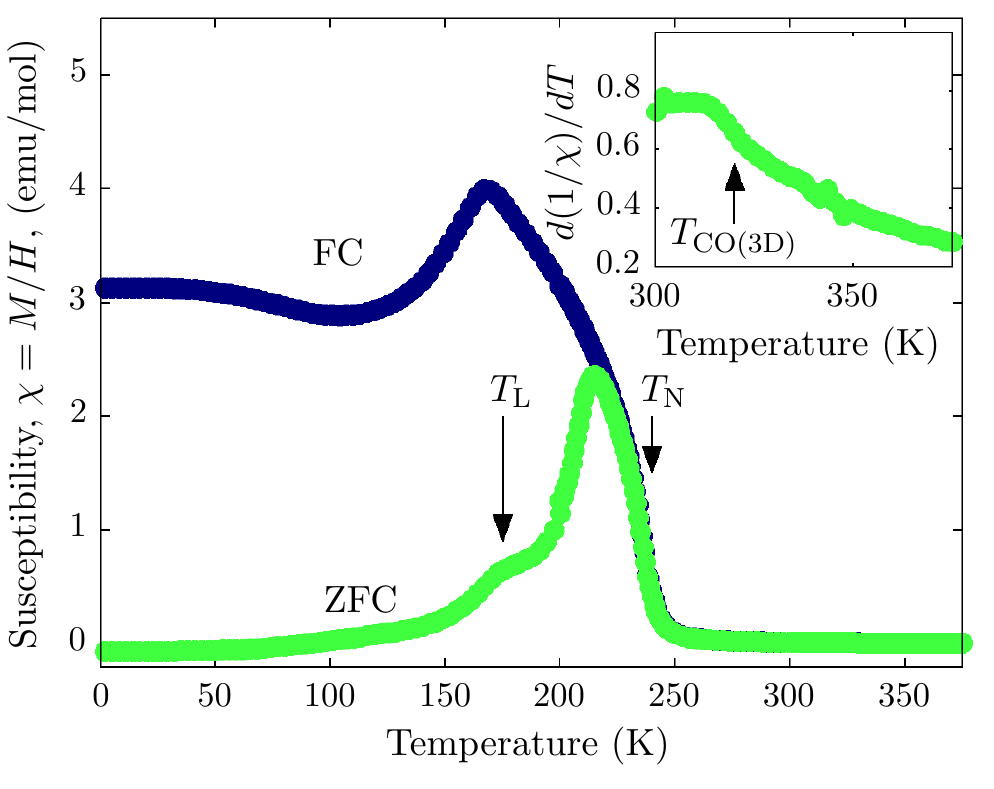}
\caption{\label{fig:LFO_SQUID_data}(Color online) Susceptibility of Sample 2 measured with a magnetic field of $H=1000$\,Oe applied parallel to the $c$ axis. The green (blue) points denote the ZFC (FC) data. The insert shows the derivative of inverse ZFC susceptibility plotted against temperature and highlights the 3D charge ordering transition at $\sim320$\,K.}
\end{figure}

Figure~\ref{fig:LFO_SQUID_data} shows the field-cooled (FC) and zero-field-cooled (ZFC) susceptibilities of Sample~2, measured with a field of 1000\,Oe parallel to the $c$ axis.  The data for Sample~1 are qualitatively similar. The general features of the susceptibility curves resemble those reported in numerous other \LFO\ publications.\cite{Iida1993, Park2007, Zhang2007, Wu2008, Phan2010, Wang2010, deGroot2012Jan} The sharp rise in susceptibility at $T_{\rm N} \approx 240$\,K is caused by the ferrimagnetic ordering transition, and a second transition is observed at $T_{\rm L} \approx 170$\,K. The latter is reported to be associated with a monoclinic distortion which is accompanied by changes in the magnetic structure.\cite{Christianson2008,Xu2008}

Previous studies have shown that the magnetic properties of \LFO\ are dependent on the precise oxygen content of the samples. The FC and ZFC susceptibilities in Fig.~\ref{fig:LFO_SQUID_data} are consistent with those reported for \LFO\ with a slight oxygen excess.\cite{Wang2013} Similar data are shown in Refs.~\onlinecite{Iida1993, Wu2008} which report quasi-2D -- but not 3D -- magnetic order. The susceptibility of a nominally stoichiometric crystal which exhibits 3D magnetic order between $T_{\rm L}$ and $T_{\rm N}$ is reported in Ref.~\onlinecite{Christianson2008}. Excess oxygen is likely to modify the weak inter-bilayer magnetic coupling and reduce the degree of order in the out-of-plane direction. The magnetic excitations in the energy range probed here are not sensitive to inter-bilayer coupling, and so not expected to depend on whether the magnetic order is 3D or 2D.

\subsection{Magnetic Excitation Spectrum}

\begin{figure}
\centering
\includegraphics[width=\columnwidth]{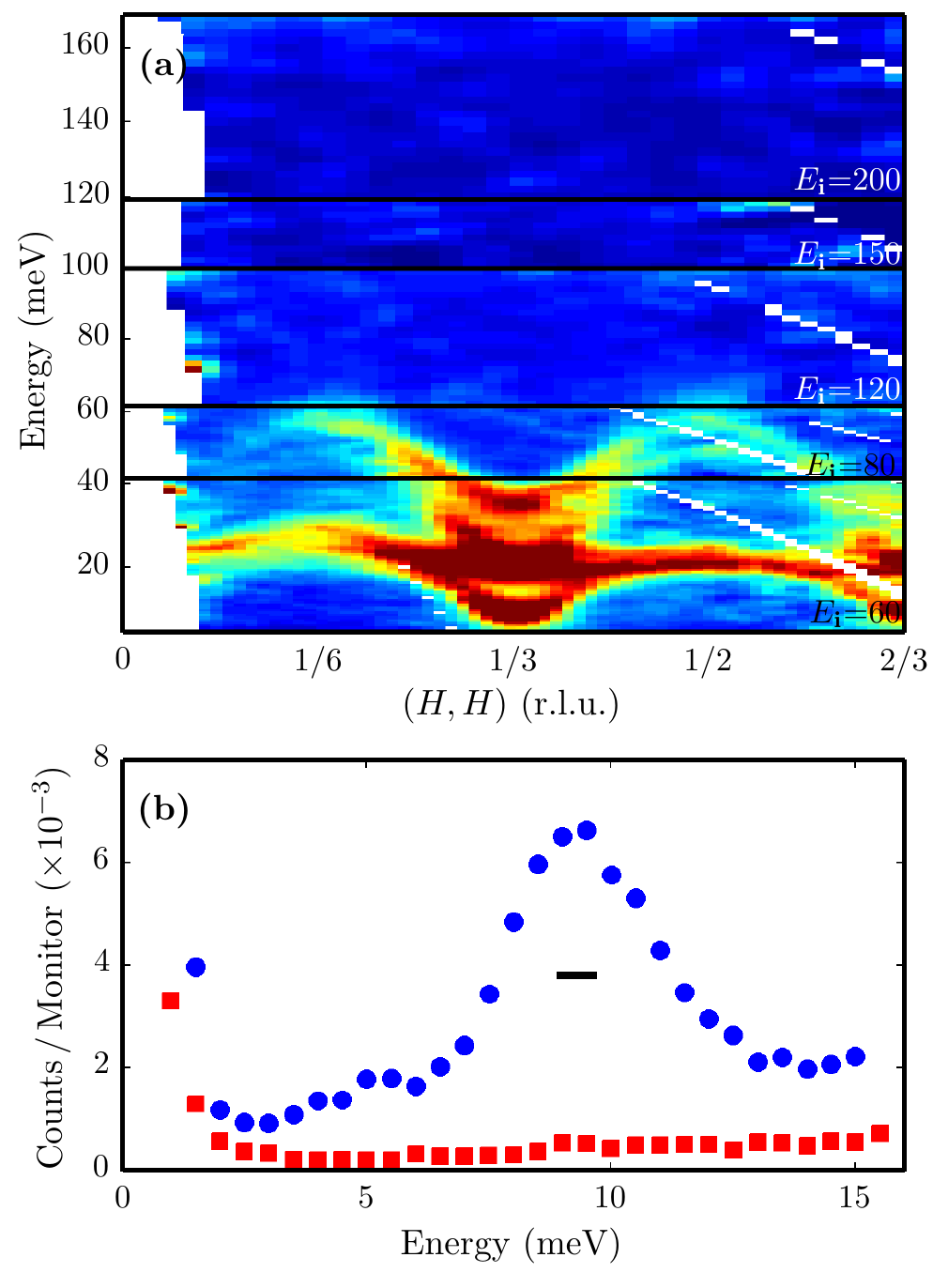}
\caption{\label{fig:LFO_INS_data_overview}(Color online) Inelastic neutron scattering from \LFO. (a) Overview of MAPS measurements on Sample 2 as a function of wave vector along the $(H,H)$ direction. The different panels contain data recorded with incident energies $\Ei=60$, 80, 120, 150 and 200\,meV (labelled). The intensity has been scaled independently in each panel for clarity. (b) Energy scans centered at fixed wave vectors $\Gamma=(1/3,\,1/3,0)$ (blue circles) and K$=(0,\,2/3,0)$ (red squares) measured on IN8. Both scans were measured with fixed $k_{\rm f}=2.662$\,\invAA. In this experimental setup, 1000 monitor corresponds to a counting time of $50-100$\,s depending on $E$. Statistical errors are smaller than the marker size. Data from the K point provide an estimate of the non-magnetic background at low energies. The horizontal bar represents the energy resolution (FWHM) estimated from a simulation of the energy scan.\cite{Restrax} }
\end{figure}

Figure~\ref{fig:LFO_INS_data_overview} provides an overview of the magnon spectrum of \LFO. Figure~\ref{fig:LFO_INS_data_overview}\col{(a)} shows the spectrum measured by TOF neutron scattering from Sample 2 along the $(H,H)$ direction --- see Fig.~\ref{fig:LFO_recip_latt}(a). The intensity map is a composite of data recorded with several different energies. The data extend up to 170\,meV, but no features could be observed above $60$\,meV. Below $60$\,meV, a series of magnon bands can be seen. The modes appear broader than expected from the instrumental resolution and overlap at various points in the spectrum. A significant structure factor modulation in the intensity is evident from the asymmetry of the intensity either side of the $\Gamma=(1/3,\,1/3)$ supercell zone centre position, most noticeably in the modes below 40\,meV at the equivalent supercell zone boundary positions M$=(1/6,\,1/6)$ and $(1/2,\,1/2)$.

Figure~\ref{fig:LFO_INS_data_overview}\col{(b)} shows an energy scan at the $\Gamma=(1/3,\,1/3,0)$ zone centre measured on Sample 1 with the IN8 TAS. In this low energy part of the spectrum there is a peak centred at 9.4\,meV, corresponding to the minimum in the dispersion of the lowest energy magnon mode. After taking into account the experimental resolution this peak is consistent with the Ising gap of $\sim 8$\,meV reported in a powdered \LFO\ sample.\cite{Bourgeois2012Jul} However, some magnetic intensity above background can also be seen below the gap. This low energy signal cannot be explained by the tail of the $(1/3,1/3,0)$ magnetic Bragg peak, which does not extend above 3\,meV. We shall discuss the possible origin of this below-gap scattering in Section~\ref{sec:Analysis and Discussion}.

\begin{figure}
\centering
\includegraphics[width=\columnwidth]{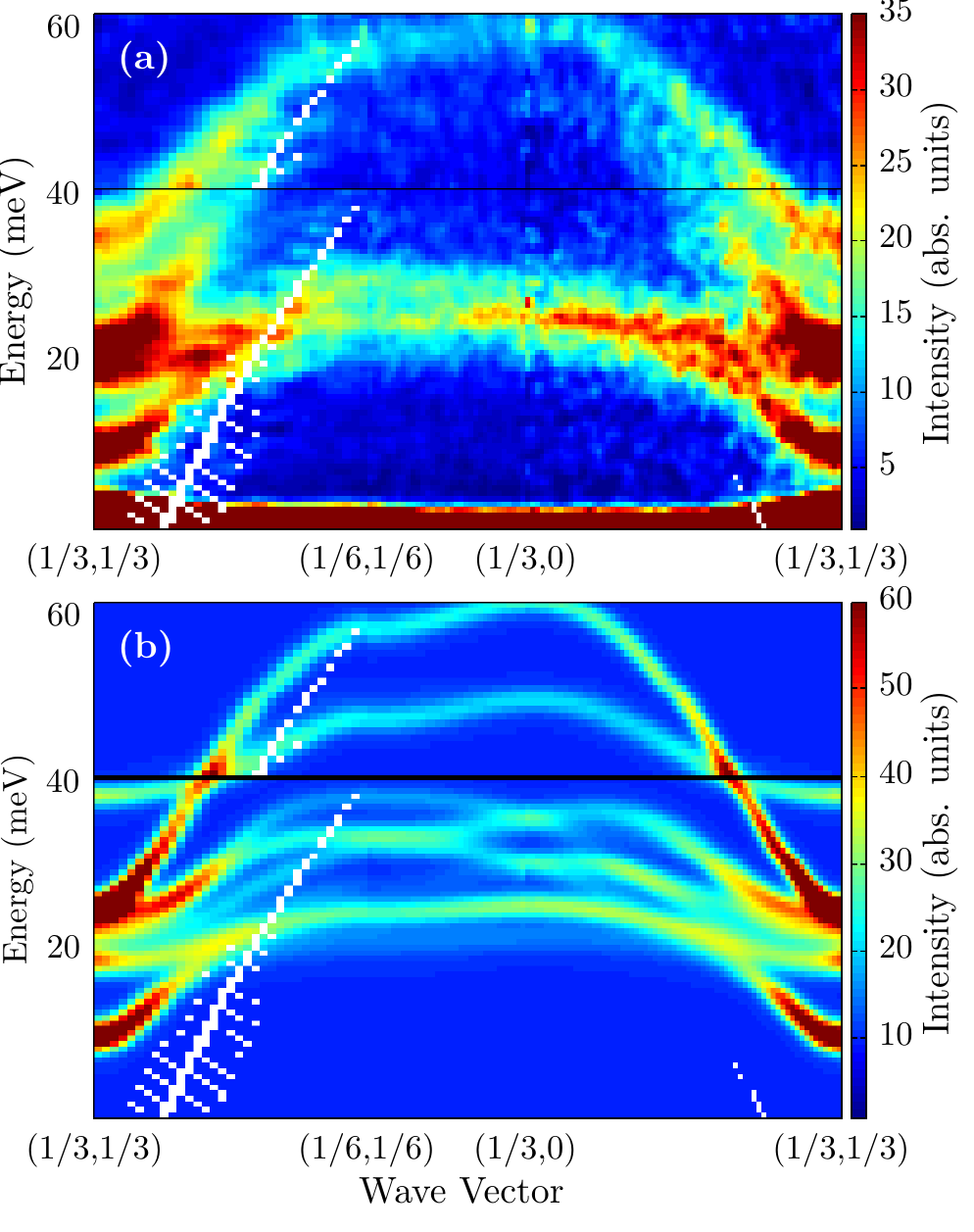}
\caption{\label{fig:LFO_INS_spectrum_detailed}(Color online) Observed and simulated magnon spectra of \LFO. (a) Spectrum along the $\Gamma\rightarrow$\,M\,$\rightarrow$\,K\,$\rightarrow\Gamma$ path in reciprocal space. The data are for Sample 2, and were measured on the MAPS spectrometer with $E_{\rm i}=60$\,meV (lower part) and $E_{\rm i}=80$\,meV (upper part). Intensities are in absolute units of mb\,sr$^{-1}$\,meV$^{-1}$\,f.u.$^{-1}$, where f.u. stands for ``formula unit'' of \LFO. (b) The simulated spectrum from the spin-wave model with best-fit parameters (see main text). The prefactors contained in Eq.~\ref{eq:S(Q,E)} together with $g_{\alpha}=2$ in Eq.~\ref{eq:Saa} have been included, so that the intensity is in the same absolute units as the data in (a).}
\end{figure}

Measurements of the magnon spectrum along the $\Gamma\rightarrow {\rm M}\rightarrow {\rm K}\rightarrow\Gamma$ path in reciprocal space [see Fig.~\ref{fig:LFO_recip_latt}(a)] are displayed in more detail in Fig.~\ref{fig:LFO_INS_spectrum_detailed}(a). Four modes can be resolved clearly in this color intensity map. Evidence for a further two modes is found from a more quantitative analysis of the full spectrum, as described below.

\begin{figure}
\centering
\includegraphics[width=\columnwidth]{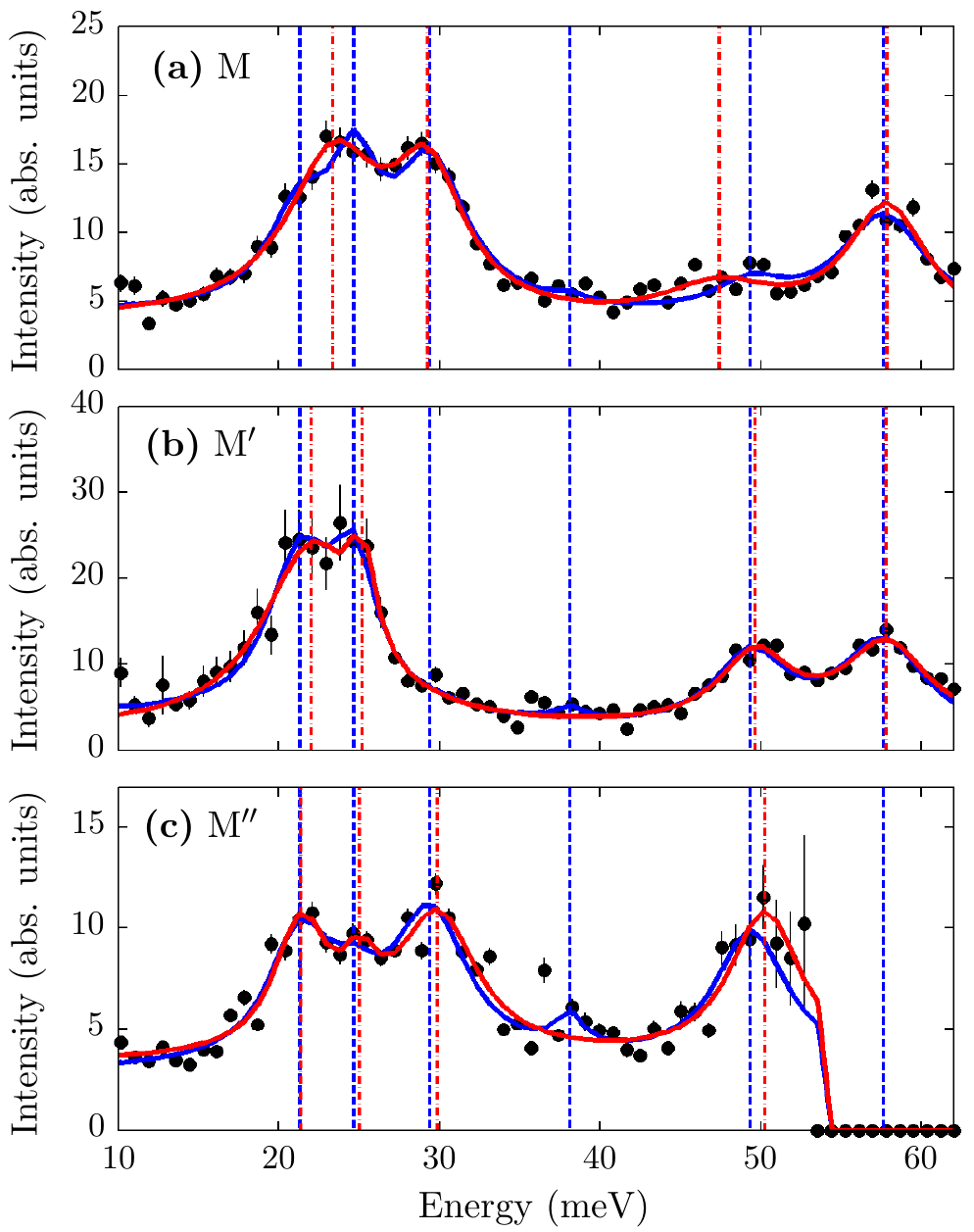}
\caption{\label{fig:fitted_Ecuts}(Color online) Energy cuts at the (a) ${\rm M}=(1/6,1/6)$, (b) ${\rm M}^{\prime}=(1/6,2/3)$, and (c) ${\rm M}^{\prime\prime}=(2/3,1/6)$ positions. The black points are experimental data from Sample 2 measured on the MAPS spectrometer with $\Ei=80$\,meV. All intensities are in absolute units of mb\,sr$^{-1}$\,meV$^{-1}$\,f.u.$^{-1}$. The solid blue lines show the results of the semi-global fit (see main text). The red lines show the best fits to each individual energy cut. The blue (red) vertical dashed (dot-dashed) lines show the fitted positions of the peaks using the two types of fit.}
\end{figure}

Figure~\ref{fig:fitted_Ecuts} shows energy cuts at the M, M$^{\prime}$, and M$^{\prime\prime}$ wave vector positions in the Brillouin zone (BZ). These positions are located at mid-points along the sides of the 2D zone boundary --- see Fig.~\ref{fig:LFO_recip_latt}(a) --- and are symmetry-inequivalent (they are not connected by reciprocal lattice vectors of the magnetic superlattice). Hence, the modes at each inequivalent zone boundary mid-point can have different energies. Additionally, structure factor effects can create significant variations between the intensities at the different positions. Both of these effects can be seen to some extent in the energy cuts in Fig.~\ref{fig:fitted_Ecuts}.  There are intensity differences between modes that are common to all the cuts, such as those near $30$ and 50\,meV, as well as small but resolvable differences in the mode energies as determined from fits to a series of Lorentzian peak functions (shown as solid red lines). When data along all wave vector paths are examined, we find evidence for a maximum of six modes in the spectrum, as can be seen in Fig.~\ref{fig:fitted_Ecuts}.

\section{\label{sec:Analysis and Discussion}Analysis and Discussion}

\subsection{Magnetic Ground State}

As outlined earlier, the two most likely models for the charge and magnetic structures in \LFO\ have bilayers formed from A and B monolayers in the sequence either AB-BA or AA-BB, with parallel alignment of the fM monolayers in the bilayers. The measured dispersion shows evidence for six magnon bands. This indicates that a bilayer magnetic unit cell containing six spins describes the majority of the sample, which is consistent with the AB-bilayer model shown in Figs.~\ref{fig:LFO_charge_mag_struct}\col{(b)} and \ref{fig:LFO_3D_struct_comparison}\col{(a)}. The AA-BB bilayer model depicted in Fig.~\ref{fig:LFO_3D_struct_comparison}\col{(b)} has two distinct bilayer magnetic structures (AA and BB) containing six spins each, and so the spectrum will contain twelve magnon bands.\footnote{Note, however, that in the absence of inter-monolayer interactions both the AB-BA and AA-BB bilayer structures would give the same spectrum, a superposition of the spectra of the uncoupled A and B monolayers. An individual A or B monolayer has a magnetic unit cell containing three spins, giving three magnon bands. The magnon spectrum of a system of uncoupled A and B monolayers would therefore contain a total of six bands.} It is very unlikely that the two sets of six magnon bands arising from the individual AA and BB bilayers are degenerate, since the magnetic anisotropy and exchange interactions within AA and BB layers will be very different. In fact, the most prominent excitations are observed to have a gap of about 8\,meV at the $\Gamma$ position, see Fig.~\ref{fig:LFO_INS_data_overview}(b), whereas magnons in a BB-bilayer comprising two \Fethree-rich monolayers are expected to be more weakly gapped because \Fethree\ has much smaller single-ion anisotropy than \Fetwo. As we do not observe twelve bands of similar intensity or two gaps, a majority AA-BB ordering is not compatible with our data. However, a minority of AA-BB order could provide an explanation for the weak magnetic intensity observed below 8\,meV --- Fig.~\ref{fig:LFO_INS_data_overview}(b). The mixed phase state proposed in Ref.~\onlinecite{Bourgeois2012Jul} with roughly equal numbers of FM and AFM bilayers would also lead to more than six bands, and so is also not compatible with our data.

The discrepancy between the AB-bilayer model that seems to account for the magnon spectrum and the AA-BB model deduced from diffraction data is puzzling. However, we note that the latter model was derived from data collected at 210\,K, whereas our neutron inelastic scattering measurements were performed at low temperatures. It is possible that the magnetic structure changes between these two very different temperatures. Indeed, Christianson {\it et al.} reported significant changes in the magnetic diffraction peak intensities on cooling through the $T_{\rm L}$ transition.\cite{Christianson2008} This observation could provide a reconciliation between the two models. Another possibility is that small differences in oxygen content could lead to different magnetic structures within the bilayers.



Based on these considerations, we shall proceed on the assumption that the in-plane magnetic order in our samples is the FM-coupled AB-bilayer order proposed in Refs.~\onlinecite{Ko2009, Kuepper2009}. This is consistent with the six magnetic modes measured, the single prominent excitation gap, and reflects the similarity between the measured susceptibility of Sample 2 (Fig.~\ref{fig:LFO_SQUID_data}) and the sample of Wu, \etal\cite{Wu2008} (subsequently used in Ref.~\onlinecite{Ko2009}). This bilayer charge and magnetic structure is depicted in Fig.~\ref{fig:LFO_charge_mag_struct}\col{(b)}. It can be seen from the figure, that the charge and magnetic order of the supercell breaks the local $\bar{3}$m symmetry characteristic of the crystallographic unit cell. This leads to six domains contributing to the measured magnon spectrum. The effect of multiple domains has been accommodated for in the analysis of the spectrum, as described below and in Appendix~\ref{app:Domains}.

\subsection{\label{sec:Spin Wave Model}Spin Wave Model}

We develop a linear spin-wave model (SWM) to describe the magnetic spectrum. The SWM is derived from a Heisenberg Hamiltonian with uniaxial anisotropy,
\begin{equation}
\mathcal{H}=\sum_{\langle i,j\rangle}J_{ij}\bm{S}_{i}\cdot\bm{S}_{j}-\sum_{i}\mathcal{D}_{i}(S_{i}^{z})^2,
\label{eq:spin hamiltonian}
\end{equation}
where $J_{ij}$ are the exchange parameters, and $\mathcal{D}_{i}$ are single-ion anisotropy terms defined separately for \Fetwo\ and \Fethree\ ions. The summation is over all pairs of spins, with each pair counted once, denoted by $\langle i,j\rangle$. Figure~\ref{fig:LFO_charge_mag_struct}\col{(b)} specifies the exchange interactions used in the model. The different electronic structures of \Fetwo\ and \Fethree\ mean that the superexchange pathways between each ion will be distinct. Therefore, the exchange parameters represent the \Fetwo--\Fethree\ coupling within each monolayer ($J_{\rm{A1}}$ and $J_{\rm{B1}}$), the \Fetwo--\Fetwo\ coupling on the \Fetwo-rich monolayer ($J_{\rm{A2}}$), the \Fethree--\Fethree\ coupling on the \Fethree-rich monolayers ($J_{\rm{B2}}$), and a single inter-monolayer exchange ($J_{\rm{AB}}$). This defines a minimal model which includes only nearest-neighbour (nn) interactions within and between the monolayers.\footnote{When considering the full symmetry of the supercell, there are a total of 15 inequivalent nn superexchange pathways.\cite{Xiang2009} However, when the different oxygen environments of each Fe site are assumed to be equivalent this contracts to the five defined and maintains the simplicity of the model.} The exchange interactions are expected to be short range since \LFO\ is an insulator. This assumption is supported by the DFT calculations of Xiang \etal\cite{Xiang2009} in which next-nearest-neighbour interactions were found to be negligible. The separate single-ion anisotropy terms describe the Ising-like anisotropy in \LFO, while allowing the strength of the anisotropy to be different for the \Fetwo\ and \Fethree\ ions.

During the analysis, the proposed Hamiltonian was checked to ensure that it gave the assumed magnetic ground state. This was done via a mean field calculation, based on the simplification of the Hamiltonian in Eq.~\ref{eq:spin hamiltonian} to the Ising limit:
\begin{equation}
\mathcal{H}_{\rm{Ising}}=\sum_{\langle i,j\rangle}J_{ij}S^{z}_{i} S^{z}_{j}.
\label{eq:mf hamiltonian}
\end{equation}
With six Fe ions in the magnetic unit cell, $2^6 = 64$ Ising spin states exist. For a given set of $J_{ij}$, the ground state spin structure is the one which minimises $\mathcal{H}_{\rm{Ising}}$.

By considering all the exchange pathways defined in Fig.~\ref{fig:LFO_charge_mag_struct}\col{(b)}, the ground state Ising spin structure can be identified for any given values of $J_{ij}$. Performing this calculation for a variety of exchange parameters reveals the accepted spin structure as the lowest energy ordering. Therefore, the chosen Hamiltonian is consistent with the proposed magnetic order for suitable choices of $J_{ij}$ when there is strong $c$ axis anisotropy.

\subsection{Analysis of the magnon dispersion}

We extracted the magnon dispersion by fitting a series of Lorentzian peaks to energy cuts (\E-cuts) at positions across the BZ. A linear background was included in all fits. To extract the mode energies from \E-cuts we initially performed a `semi-global' fit. By this we mean that a single set of mode energies and widths were fitted to all \E-cuts at equivalent or approximately equivalent points in the BZ. For example, the ${\rm M}$, ${\rm M}^{\prime}$ and ${\rm M}^{\prime\prime}$ points are not equivalent for the bilayer magnetic structure but are equivalent for the magnetic structure of the individual A and B monolayers, since the magnon spectrum of the monolayers has 6-fold symmetry in reciprocal space (see Appendix~\ref{app:Domains}). Therefore, when the inter-layer coupling is weak the energies of the modes at ${\rm M}$, ${\rm M}^{\prime}$ and ${\rm M}^{\prime\prime}$ are approximately the same. The amplitudes of the peaks were allowed to vary separately in each line cut since the structure factors are different at inequivalent positions. In a second iteration, we fitted each individual \E-cut separately, and did not constrain the energies at approximately equivalent {\bf Q} positions to be equal. Examples of both types of fits are shown in Fig.~\ref{fig:fitted_Ecuts}.

\begin{figure}
\includegraphics[width=\columnwidth]{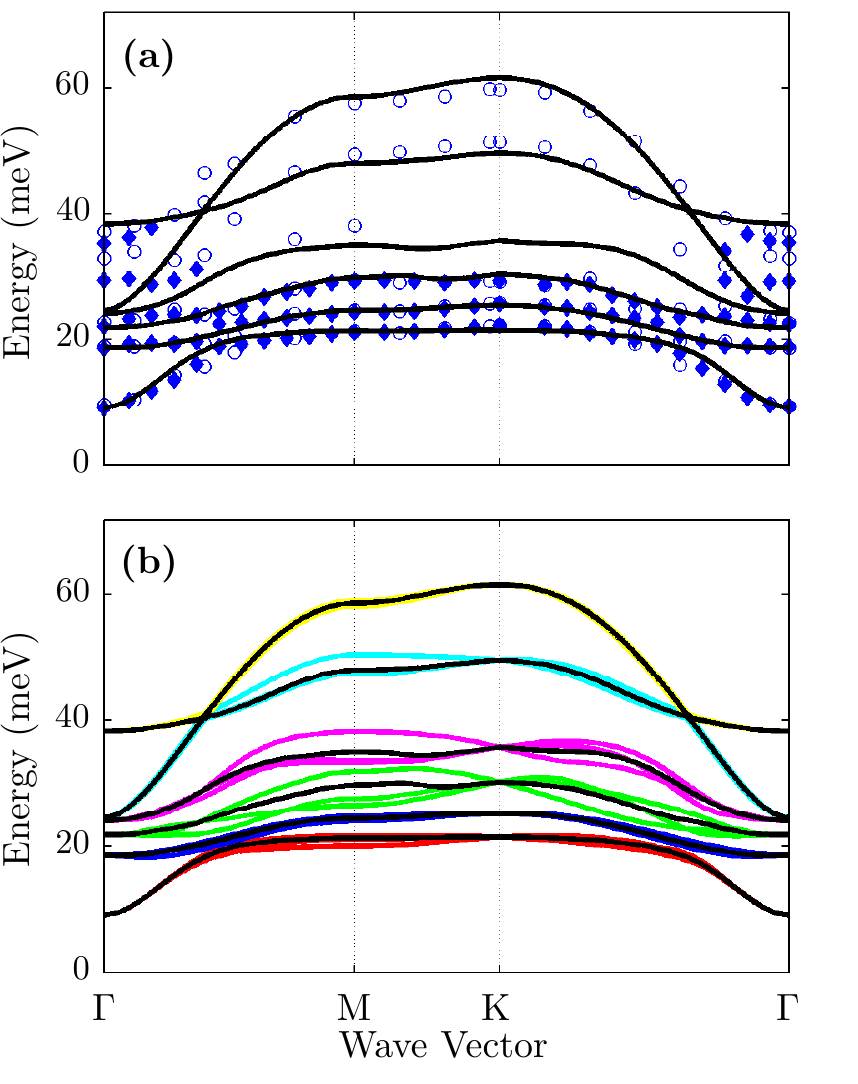}
\caption{\label{fig:dispersion fitting results}(Color online) Measured and refined magnon dispersion relation of \LFO. (a) Blue filled diamond (open circle) data points show mode energies found from semi-global fits to the $\Ei=60$\,meV (80\,meV) neutron spectra. The semi-global fitting method (see text) means that observed dispersion relations are an average over the paths $\Gamma \rightarrow {\rm M} \rightarrow {\rm K} \rightarrow \Gamma$, $\Gamma \rightarrow {\rm M^{\prime}} \rightarrow {\rm K} \rightarrow \Gamma$ and $\Gamma \rightarrow {\rm M^{\prime\prime}} \rightarrow {\rm K} \rightarrow \Gamma$.  (b) The fitted dispersions arising from all domains expected in the magnetic unit cell depicted in Figure~\ref{fig:LFO_charge_mag_struct}\col{(b)}. Each of the six magnetic domains has a set of 6 bands, which are shown with different colored lines. The intensity-weighted domain average of each mode (Appendix~\ref{app:Domains}) is plotted as a solid black line in both panels.}
\end{figure}

The semi-global fit made it possible to obtain the energies of six modes with small experimental uncertainties across much of the BZ. Figure~\ref{fig:dispersion fitting results}(a) shows the fitted dispersion relations along the high symmetry path $\Gamma \rightarrow {\rm M} \rightarrow {\rm K} \rightarrow \Gamma$. Fitted points from both the $\Ei=60$ and $80$\,meV spectra are included. There is good agreement between the $\Ei=60$ and $80$\,meV points below 30\,meV. However, at higher energies around the $\Gamma$ position there are small discrepancies which introduce some uncertainty in the dispersion in this region. A numerical implementation of linear spin wave theory\cite{spinwave_dispersions} was used to refine the dispersion of the minimal model against the data.  To accommodate for the effect of the coexisting magnetic domains, the calculated dispersion used in the fit was a weighted average over the six domains with weighting proportional to the calculated intensities, see Appendix~\ref{app:Domains}.

A systematic search was performed to find the best-fit model.  Each of the $J$ parameters was first coarsely varied between -5 to 5\,meV in 2.5\,meV steps, giving an array of $5^5$ starting parameter sets. Then, they were varied more finely between 0 to 5\,meV in 0.5\,meV steps giving $11^5$ starting parameter sets.  The fine grid was limited to AFM interactions ($J\geq0$) because most interactions are predicted to be AFM.\cite{Iida1993, Xiang2009, Ko2009}

The search was guided by the mean-field model defined by Eq.~\ref{eq:mf hamiltonian}. The mean-field energies of all distinct Ising spin arrangments were calculated for each parameter set. Starting parameters which energetically favor a ground state spin structure other than that depicted in Fig.~\ref{fig:LFO_charge_mag_struct}\col{(b)} were discarded. The remainder formed a list of 201 and 5329 sets of starting parameters for the coarse and fine grids, respectively. The starting values of the anisotropy parameters were set to 3\,meV for both the \Fetwo\ and \Fethree\ ions, even though \Fetwo\ has a much greater single-ion anisotropy than \Fethree. This was to avoid any unintentional bias in the fitting, and to check that the fit could distinguish the \Fetwo\ and \Fethree\ sites on its own.

Fits were then performed using all remaining sets of starting parameters. In a few places several extracted mode energies are in close proximity, such as between $15<E<40$\,meV around the $\Gamma$ position --- see Fig.~\ref{fig:dispersion fitting results}(a). In these cases two calculated modes were refined against a single data point. Each data point was weighted by the inverse square-root of its error (i.e.\, $ w_{j}=1/\sqrt{\delta E_j}$ is the weight of the $j^{\rm{th}}$ data point). This fitting produced $\sim40$ distinct converged solutions. In the final step, these converged solutions were used to provide starting parameters for fits to the dispersion data extracted from $E$-cuts at individual \Q\ positions, which in contrast to the semi-global fit does not assume inequivalent points in the BZ to be equivalent (but does include domain-averaging).


The dispersion relations from the best fit, i.e.~that with the lowest $\chisq$ value, are shown in Fig.~\ref{fig:dispersion fitting results}, and the fitted parameters are listed in Table~\ref{tab:LFO_LSWM_Js}. This solution was found to be stable, in the sense that it was consistently found when the data were re-fitted using starting parameters randomly shifted a small amount away from the converged values.  The dispersion relations calculated from the best fit are compared in Fig.~\ref{fig:dispersion fitting results}(a) with the data obtained from the semi-global fits. The agreement is seen to be very good.

To evaluate the influence of the magnetic domains, we show in Fig.~\ref{fig:dispersion fitting results}(b) the dispersion relations calculated separately for each of the magnetic domains. Equivalent modes from different domains are shown in the same color. The inversion symmetry of reciprocal space means that the six magnetic domains result in three distinct magnon dispersions. Each dispersion has six modes, giving 18 distinct modes in total.  At the $\Gamma$ and K points the set of mode energies for each domain are the same by symmetry. Elsewhere, equivalent modes are split by up to 5\,meV, though usually much less than that.  Also shown in Fig.~\ref{fig:dispersion fitting results}(b) are the domain-averaged dispersion relations used in the refinement of the SWM. We obtained these by averaging the energies of equivalent modes from different domains assuming an equal population of domains and using the calculated magnon scattering intensities as weighting factors. Figure~\ref{fig:dispersion fitting results}(b) shows that to within the energy resolution of our measurement, typically 3--4\,meV, the equivalent modes from different domains cannot be resolved over much of the spectrum. This explains why only six modes can be seen in the experimental data, and justifies the use of the semi-global fitting method.

\begin{table}[t]
\centering
\renewcommand{\arraystretch}{1.5}
	\begin{tabular}{c c c c}
	\hline\hline
	Parameter & Best Fit & 2$^{\rm nd}$ Best Fit & DFT\cite{Xiang2009} \\
	\hline
	$J_{\rm A1}$ & 1.185(4) & 1.184(5) & 4.0 \\
	$J_{\rm A2}$ & 0.935(8) & 1.356(6) & 2.75 \\
	$J_{\rm B1}$ &  0.868(5) & -0.08(6) & 1.87 \\
	$J_{\rm B2}$ & 3.407(8) & 3.405(8) & 7.3 \\
	$J_{\rm AB}$ & -1.098(8) & -1.05(1) &  0.9 \\
	$\mathcal{D}_{\rm Fe^{2+}}$ & 5.53(2) & 6.02(2) & -- \\
	$\mathcal{D}_{\rm Fe^{3+}}$ & 0.60(1) & 0.84(2) & --\\
	\chisq\ & 6.49 & 6.55 & -- \\
	\hline\hline
	\end{tabular}
\caption{\label{tab:LFO_LSWM_Js}Exchange and anisotropy parameters for \LFO. Refined parameters from the best and second best fits to the magnon dispersion are listed together with their \chisq\ goodness-of-fit values. The exchange parameters are defined in Fig.~\ref{fig:LFO_charge_mag_struct}\col{(b)}. The errors on the fitted parameters are the standard deviations found from the fitting procedure. The results of \textit{ab initio} density function theory (DFT) calculations from Ref.~\onlinecite{Xiang2009} are presented alongside for comparison. The DFT values quoted in the table are an average of the nn parameters that become equivalent in the minimal model used in our study.}
\end{table}

The second-best converged solution is characterised by a $\chisq$ value ($\chisq=6.55$) only slightly larger than that of the best fit ($\chisq=6.49$). For comparison, the fitted parameters for the second-best fit are also listed in Table~\ref{tab:LFO_LSWM_Js}. Although some parameters are very similar to those of the best-fit solution, others are not, most noticeably $J_{\rm A2}$ and $J_{\rm B1}$. These parameters lead to a dispersion that fails to reproduce the accurately measured dispersion of the lowest energy mode around the BZ center. Therefore, although the values of \chisq\ for the best and second-best fits are not very different, a more sophisticated weighting scheme which better reflected the experimental precision of the low energy data near $\Gamma$ would differentiate these fits more clearly. Other converged solutions have significantly higher \chisq\ values ($\ge8.30$) in addition to noticeably worse qualitative agreement with the measured dispersion.

The best-fit solution provides a very good description of the measured dispersion of \LFO. Furthermore, the converged parameters successfully recover the expected strong and weak anisotropies of the \Fetwo\ and \Fethree\ ions, respectively. The exchange parameters are found to be principally AFM, as expected from previous studies.\cite{Iida1993, Xiang2009, Ko2009} Additionally, a comparison of our fitted parameters to the equivalent parameters found from \textit{ab initio} calculations\cite{Xiang2009} reveals differences in magnitude but many qualitative similarities. This comparison is presented in Table~\ref{tab:LFO_LSWM_Js}, which lists the results of Ref.~\onlinecite{Xiang2009}. To enable this comparison, the full set of 15 nn exchange parameters considered in Ref.~\onlinecite{Xiang2009} were grouped into those with approximately equivalent paths and averaged to obtain the reduced set of 5 parameters used in our minimal SWM. Both \textit{ab initio} and SWM parameters show that $J_{\rm B2}$ is the strongest and $J_{\rm A1}$ is the next-strongest interaction.

\subsection{Simulated Spectrum}

Figure~\ref{fig:LFO_INS_spectrum_detailed}\col{(b)} shows the magnon spectrum for the best-fit model calculated from Eqs.~\ref{eq:S(Q,E)} and~\ref{eq:Saa}. The calculation is a simulation of the intensity map in Fig.~\ref{fig:LFO_INS_spectrum_detailed}\col{(a)} and takes into account the variation of $\bf Q$ with $E$ intrinsic to the TOF neutron scattering spectra. The energy $\delta$-function in Eq.~\ref{eq:S(Q,E)} is replaced by a Gaussian with an energy-dependent width to simulate the spectrometer resolution, and the $g$-tensor is taken to be isotropic with diagonal elements equal to 2 to put the intensities approximately on an absolute scale. We have not included the effects of neutron absorption and self-shielding, which we estimate will reduce the intensities by about $15-20$\% depending on $E$. The intensity was calculated for all six magnetic domains and averaged to arrive at the presented spectrum.


Overall, the simulation reproduces the intensity variation in the data very well. In particular, we do not see additional modes that are not described by the model. If there are more modes in the spectrum then they are too weak to detect. Conversely, all modes shown by the simulation are also visible in the data, although we note that the mode that disperses between 40 and 50\,meV is very weak in the vicinity of the M point in Fig.~\ref{fig:LFO_INS_spectrum_detailed}(a). This mode is more intense at the approximately equivalent M$^{\prime}$ and M$^{\prime\prime}$ positions --- see Figs.~\ref{fig:fitted_Ecuts}(b) and (c) --- which is how we were able to determine its dispersion shown in Fig~\ref{fig:dispersion fitting results} accurately.

 The simulation highlights two other features worth mentioning. First, there is a discrepancy in the energy widths of the magnon resonances. The measured widths are systematically broader than the experimental resolution. This is partly explained by the domain splitting of the modes, but additional broadening could also arise from in-plane defects in the magnetic order. Second, there is an ambiguity about whether the highest and next-highest magnon bands cross near $\Gamma$.
To the naked eye, the intensity map in Fig.~\ref{fig:LFO_INS_spectrum_detailed}(a) would suggest no band crossing, whereas our best-fit model does have a crossing, Fig.~\ref{fig:LFO_INS_spectrum_detailed}(b). The magnon energies extracted from fits to the measured spectra do not provide a conclusive dispersion of the bands in this region --- see Fig.~\ref{fig:dispersion fitting results}. We did find fits in which these bands do not cross, but these fits gave a poorer description of other parts of the spectrum. Since our fitting procedure was rather exhaustive we are confident that any remaining discrepancies of this nature cannot be resolved within the constraints of our minimal model.

We emphasize that the model we have used here is based on a spin-only Hamiltonian with a minimum set of nn exchange interactions. Inclusion of the full set of distinct nn exchange paths or more distant neighbours might improve the model but would be impractical to fit given the large number of parameters and the long computation times associated with numerically-implemented spin wave theory.  We have neglected effects due to spin--orbit coupling other than the Ising-like single-ion anisotropy. The introduction of an anisotopic $g$-factor, for example, would have some influence on the intensities of the modes. We have also neglected phonon scattering in the experimental data. This is reasonable because at the small $Q$ values of our data the strong magnon scattering from the large Fe moments is expected to be much greater than the phonon scattering. It is possible, however, that some of the discrepancies around 35\,meV found in our fits to data near the $\Gamma$-point are due to optic phonons.

\section{Conclusions}

In this work we have developed a simple spin model of \LFO\ through a comprehensive analysis of inelastic neutron scattering spectra measured throughout the 2D Brillouin zone. Our results are consistent with a predominant spin and charge-ordered state of \LFO\ at low temperatures consisting of polar bilayers built from one \Fetwo-rich and one \Fethree-rich monolayer, with a parallel alignment of the ferrimagnetic moment on each monolayer. This is in agreement with some but not all previous experiments. Notably, our findings are not consistent with the charge segregated AA-BB bilayer ordering proposed to explain diffraction results.\cite{deGroot2012Jan}

Our study demonstrates how measurements of excitation spectra can provide stringent constraints on the nature of the ground state in magnetic systems, and provides an important basis for the analysis of future experiments on \LFO\ designed to probe the precise bilayer stacking and to discriminate ferroelectric and antiferroelectric ordering at low temperatures.

\begin{acknowledgments}
We wish to thank R. Coldea and M. Rotter for help with the numerical solution of the linear spin-wave model, and R. D. Johnson for discussions. This work was supported by the UK Engineering \& Physical Sciences Research Council.
\end{acknowledgments}

\appendix
\section{\label{app:Domains}Domains in Magnetically Ordered AB-Bilayers}

The spin and charge order of an AB-bilayer shown in Fig.~\ref{fig:LFO_charge_mag_struct} breaks the three-fold rotational and mirror symmetry that the bilayer has in the absence of spin and charge order.  Assuming a purely 2D magnetic order, we expect there to be six domains in the low-temperature ordered phase.  Since reciprocal space has inversion symmetry, the six domains result in three distinct dispersion relations, each dispersion relation being found in two domains. Therefore, in a multi-domain sample the magnon spectrum measured along $\Gamma\rightarrow{\rm M}$ will be a superposition of the spectra from the $\Gamma\rightarrow{\rm M}$, $\Gamma\rightarrow{\rm M^{\prime}}$ and $\Gamma\rightarrow{\rm M^{\prime\prime}}$ directions.

The magnetic order on the individual A- and B-monolayers has 6- and 3-fold rotational symmetry, respectively. Therefore, if the monolayers were uncoupled (i.e.\ $J_{\rm{AB}}=0$), the magnon dispersion for both A- and B-monolayers would have 6-fold rotational symmetry, and positions like M, ${\rm M^{\prime}}$ and ${\rm M^{\prime\prime}}$ would be equivalent. Estimates of the inter-monolayer exchange $J_{\rm{AB}}$ suggest that it is small,\cite{Xiang2009} and so the domain splitting of the dispersion relations for the bilayer is also expected to be small. The inequivalent directions in reciprocal space which are related through the broken symmetry operations of the monolayers can therefore be considered as \emph{approximately equivalent}, as far as the magnon bands are concerned. This is supported by the data: no more than six modes are observed at any position in the BZ, but approximately equivalent wave vector positions show slight shifts ($\sim1-2$\,meV) in the energies of the modes, most noticeably at the M, M$^{\prime}$ and M$^{\prime\prime}$ positions as shown in Fig.~\ref{fig:fitted_Ecuts}. It should be noted that the variation of intensity due to the structure factor may also give rise to apparent shifts in peak positions when two or more peaks are close in energy. Peak shifts due to different mode energies or different structure factors at approximately equivalent wave vectors cannot be distinguished. To accommodate for both these potential effects, an intensity-weighted domain-averaged dispersion was used to compare with the measured spectrum. The results of such an averaging are illustrated in Fig.~\ref{fig:dispersion fitting results}\col{(b)}, where the dispersion relations for all the individual domains are plotted for the best-fit solution. The modes from different domains contributing to a single weighted average are plotted in the same color. The weighted domain average is plotted as a black line for each mode.


\bibliographystyle{apsrev4-1}
\bibliography{LuFe2O4_bib}

\end{document}